\documentclass[11pt]{iopart}
\usepackage{graphicx}
\usepackage{color}

\begin{document}
\title[Phase-field-crystal modeling of surface phase-transitions]{Phase-field-crystal modeling of the
(2$\times$1)-(1$\times$1) phase-transitions of Si(001) and Ge(001)
surfaces}
\author{Ye-Chuan Xu and Bang-Gui Liu}
\address{Institute of Physics, Chinese Academy of Sciences,
Beijing 100190, China} \address{Beijing National Laboratory for
Condensed Matter Physics, Beijing 100190, China}
\date{\today}

\ead{bgliu@mail.iphy.ac.cn}

\begin{abstract}
We propose a two-dimensional phase-field-crystal model for the
(2$\times$1)-(1$\times$1) phase transitions of Si(001) and Ge(001)
surfaces. The dimerization in the 2$\times$1 phase is described
with a phase-field-crystal variable which is determined by solving
an evolution equation derived from the free energy. Simulated
periodic arrays of dimerization variable is consistent with
scanning-tunnelling-microscopy images of the two dimerized
surfaces. Calculated temperature dependence of the dimerization
parameter indicates that normal dimers and broken ones coexist
between the temperatures describing the charactristic temperature
width of the phase-transition, $T_L$ and $T_H$, and a first-order
phase transition takes place at a temperature between them. The
dimerization over the whole temperature is determined. These
results are in agreement with experiment. This phase-field-crystal
approach is applicable to phase-transitions of other reconstructed
surface phases, especially semiconductor $n\times$1 reconstructed
surface phases.
\end{abstract}

\pacs{68.35.-p, 05.10.-a, 68.37.-d, 05.70.-a}

\maketitle

\section{Introduction}

Semiconductor surfaces are of huge importance especially in the
modern era of nanoscience and nanotechnology. Usually, a bulk
terminated surface ($1\times 1$) is unstable, will undergo a
surface reconstruction and become a stable reconstructed surface
phase ($m\times n$) \cite{monch,r001,ro}. Most of reconstructed
surfaces transit to some $1\times 1$ structures at elevated
temperatures. The Si(001) and Ge(001) reconstructed surfaces have
been extensively studied because they are closely relevant to the
modern computer technology \cite{s-si,s-ge}. For both of them, the
reconstruction is realized through the forming of regular arrays
of dimers in the top layer ($2\times 1$ or dimerized phase), which
has been confirmed by scanning-tunnelling-microscopy (STM)
experiment \cite{monch,s-si,s-ge}. It was shown experimentally
\cite{cont,break} that when heated to certain temperatures, the
$2\times 1$ reconstructed surfaces will transit to $1\times 1$
structures. There were some first-principles calculations for the
local atomic configurations of the $2\times 1$ surfaces \cite{ab},
but it is controversial even for the essence of the phase
transitions \cite{cont,break}. Considering that a structural phase
means an averaging of local atomic structures over large enough
scales, the reconstructed surface phases need further
clarification and the essential physics of the phase transitions
are still unknown. A theory that can elucidate the issues is
highly desirable.

Phase-field method is a reliable approach to modeling and
simulating structural phases and dynamical phase
transitions\cite{langer,rev}. It has been applied to various
fields such as spiral surface growth \cite{thinf}, dendritic
growth \cite{dend}, alloy solidification \cite{soli}, crystal
nucleation \cite{nucl}, step-flow growth \cite{liuf1,stepf},
epitaxial island growth \cite{lbg,rev1,rev2}, and surface phase
transition dynamics \cite{pf5}. Recently, a phase-field-crystal
approach was proposed to model the internal spatial structures of
a given phase \cite{pfc1}, and periodic lattices was obtained by
solving evolution equations derived from the free energies. This
powerful approach has been successfully used for natural modeling
of elastic interactions \cite{pfc2} and binary alloy
solidification \cite{pfc3}.

In this paper we propose a phase-field-crystal model for the
Si(001) and Ge(001) dimerized surface phases and their phase
transitions. We use a two-dimensional (2D) phase-field-crystal
variable to describe the dimerization of atoms in the top layer,
and determine the variable by solving an evolution equation
derived from the free energy. Our simulated morphology of periodic
arrays of the dimerizarion variable is in good agreement with
large-scale STM images of the Si(001) and Ge(001) surfaces in
high-quality samples \cite{r001,ro,s-si,s-ge}. Furthermore, we
derive the temperature dependence of the dimerization parameter,
and show that normal dimers and broken ones coexist between two
characteristic temperatures, $T_L$ and $T_H$, and a first-order
phase transition takes place at a phase-transition temperature
$T_c$ in between $T_L$ and $T_H$. These are in agreement with
experiment \cite{cont,break}. This phase-field-crystal approach
can be applied to other reconstructed surface phases and their
phase-transition dynamics.

The remaining part of this paper is organized as follows. In next
section we shall present our phase-field-crystal modeling for the
2$\times$1 phase and its phase transition dynamics during
transiting to the 1$\times$1 phase. In section III we shall
present our main simulated results. In section IV we shall apply
the model and the simulated results to the
(2$\times$1)-(1$\times$1) phase transitions of Si(001) and Ge(001)
dimerized surfaces. Finally, we shall give our conclusion in
section V.

\section{Phase-field-crystal modeling}

There are two kinds of the (001) surfaces, type-$S_A$ and
type-$S_B$, for both Si and Ge. They appear alternately in the
vertical direction, but do not exist in the same layer in
high-quality samples, although they can be changed into each other
by a 90$^{\circ}$ rotation \cite{s-si,s-ge}. Because the
dimerization takes place only in the top layer, it is reasonable
to describe both of the  Si(001) and Ge(001) dimerized surface
phases by a 2D model. We neglect the buckling of dimers without
losing main physics because we are mainly interested in their
high-temperature phase transitions to corresponding 1$\times$1
phases. We show the atomic configurations and corresponding
phase-field-crystal modeling in Fig. 1. We use
$\vec{r}=(\bar{x},\bar{y})$ for a 2D point, where $\bar{x}$
describes the coordinate in the dimerization direction or the
horizontal direction and $\bar{y}$ in the other direction. The key
order parameter is the dimerization-induced change of the bond
length, $\Delta b$, in the horizontal or $\bar{x}$ direction
because there is no dimerization in the other direction
\cite{cont,break}. Phase-field-crystal variable
$\phi(\bar{x},\bar{y})$ is periodic in the $\bar{x}$ direction and
uniform in the other direction. Along the $\bar{x}$ direction,
$\phi(\bar{x},\bar{y})$ reaches the maximum, $\phi_{\rm max}$, at
the center of the dimer and the minimum, $-\phi_{\rm max}$, at the
middle point of the two nearest dimers with the same $\bar{y}$.
This model is consistent with STM images of regular arrays of
dimer chains in high-quality Si(001) and Ge(001) dimerized
surfaces \cite{s-si,s-ge,cont,break}. The average value of
$\phi(\bar{x},\bar{y})$ over $\vec{r}$, $\bar{\phi}$, is set to
zero for the dimerized surfaces. The variable
$\phi(\bar{x},\bar{y})$ for the dimerized surfaces is normalized
so as to make the integration of $|\phi(\bar{x},\bar{y})|$ over
the 2D unit cell equal $2\Delta b$ because there are two top-layer
atoms in the 2D unit cell. The variable $\phi(\bar{x},\bar{y})$ is
identical to zero for the undimerized surfaces. These are enough
for the following phase-field simulation, but, with all these in
our mind, we can imagine that the morphology of the phase-field
variable $\phi(\bar{x},\bar{y})$ can be described approximately by
a simple function $\frac{\pi\Delta b}{c^2}\cos(\pi\bar{x}/c)$,
where $c$ is the lattice constant of the undimerized 2D unit cell
and the center of the dimer is taken as the zero point of
$\bar{x}$.

\begin{figure}[!tbh]
\begin{center}
\includegraphics[width=8.5cm,]{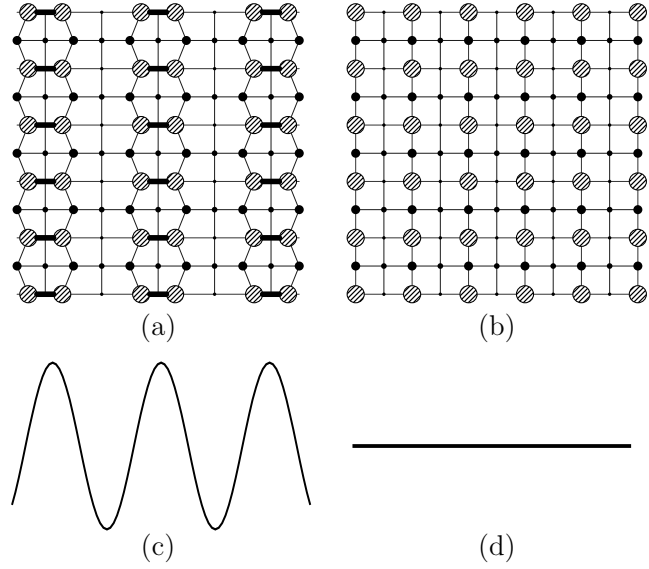}
\caption{The phase-field-crystal modeling of the Si(001) and
Ge(001) dimerized phases. (a) shows atomic models of the dimerized
(001) surfaces and (b) that of bulk-terminated (001) ones. The
dimerization happens only in the horizontal (or $\bar{x}$)
direction. The gray circle is the atom in the top layer, the
bigger black dot in the second layer, and the smaller black dot in
the third layer. The curve (c) is schematic phase-field-crystal
$\phi$ description of the dimerization-induced atomic density
change with respect to that without dimerization (d).}
\end{center}
\end{figure}

The free energy of $\phi(\bar{x},\bar{y})$ can be written as
\begin{equation}
\mathcal{F}=\lambda\int_\Omega d\vec{r}\{\frac{1}{2}\phi[A+
(q_{0}^{2}+\frac{\partial^2}{\partial
\bar{x}^2})^2]\phi-u_4\frac{\phi^4}{4}+u_6\frac{\phi^6}{6}\},
\end{equation}
where the $A$ is defined by $A=a(T-T_L)$, and the parameters
$\lambda$, $u_4$, $u_6$, and $a$ are positive constants. $q_0$ is
defined as $2\pi/\bar{c}_0$ and $\bar{c}_0$ has the meaning of
lattice constant if periodic solution is obtained. The operator
$G=(q_0^2+\frac{\partial^2}{\partial \bar{x}^2})^2$ is constructed
by fitting to the dimerization structure \cite{pfc1}. The $u_4$
and $u_6$ terms, as usual, are used to describe the first-order
phase transitions in the Si(001) and Ge(001) surfaces. The
bilinear term is sued to describe the effect of temperature and
dimerization structure. The parameter $u_6$ can be set to $1$ by
redefining the phase-field-crystal variables without losing any
physics. The evolution equation of $\phi$ is a time-dependent
Cahn-Hilliard (CH) equation \cite{ch}
\begin{equation}\frac{\partial
\phi}{\partial t}=\Gamma\nabla_r^2\frac{\delta \mathcal{F}}{\delta
\phi}+\eta , \end{equation} where $\eta$ is a Gaussian random
variable whose average value is set to zero. The first term on the
right side describes the diffusion of the phase-field variable.
Using the dimensionless phase-field variable $\psi$, dimensionless
coordinates $\vec{x}$, and dimensionless time $\tau$ defined by
\cite{pfc1}
\begin{equation}
\vec{x}=(x,y)=q_0\vec{r}, ~~\psi=\displaystyle\frac{
\sqrt[4]{u_6}}{q_0} \phi, ~~ \tau=\lambda\Gamma q_0^{6} t,
\end{equation}
we express the free energy (1) and the CH equation as
\begin{equation}
\mathcal{F}=\mathcal{F}_0\int_\Omega
d\vec{x}\{\frac{\psi}{2}[\varepsilon+(1+\frac{\partial^2}{\partial
x^2})^2]\psi-\frac{u}{4}\psi^4+\frac 16 \psi^6\}
\end{equation}
and
\begin{equation}
\frac{\partial \psi}{\partial
\tau}=\nabla^2\{[\varepsilon+(1+\frac{\partial^2}{\partial
x^2})^2]\psi-u\psi^3+\psi^5\}+\zeta
\end{equation}
Now we have only three independent parameters: $\mathcal{F}_0$,
$\varepsilon$, and $u$. $\mathcal{F}_0$ has the dimension of
energy and the other two are dimensionless. They can be expressed
in terms of original parameters: $\mathcal{F}_0=\lambda
q_0^{4}/\sqrt{u_6}$, $\varepsilon=a(T-T_L)/q_0^4=\alpha(T/T_L-1)$,
and $u=u_4/(q_0^2\sqrt{u_6}) $. The new random variable $\zeta$
whose average value is also zero is subject to the two-point
correlation function
\begin{equation}
\langle\zeta(\vec{x},\tau)\zeta(\vec{x^{\prime}},\tau^{\prime})\rangle=
D\nabla^2\delta(\vec{x}-\vec{x^{\prime}})\delta(\tau-\tau^{\prime})
\end{equation}
with $D= k_B T/ \mathcal{F}_0$. The average value of $\psi(x,y)$,
$\bar{\psi}$, is conserved by Eq. (5) and therefore $\bar{\psi}$
can be taken as an independent parameter.

\section{Main simulated results and analysis}

We solve the dimensionless evolution equation Eq. (5) by
difference method. The Laplace operator $\nabla^2$ and the
second-order differential operator $\partial^2/\partial x^2$ are
discretized by the central difference formula as usual, and the
time coordinate is discrerized using the first-order finite
differential approximation. The periodic boundary condition is
adopted for all the simulations. The initial condition is subject
to Gaussian random fluctuations. We find that resultant
equilibrium patterns are independent of the random variable
$\zeta$. We keep $\bar{\psi}=0$ because the high-temperature
1$\times$1 phase has $\psi=0$. The parameter $\mathcal{F}_0$ is
not directly relevant in our solving Eq. (5). We have tried
various values for the system sizes and the parameters $u$ and
$\varepsilon$. We choose approximately 40 periods or 80 surface
lattice constants when doing the main simulations, but confirm our
results using larger systems. We use $\delta x=\pi/16$ and
$\delta\tau=10^{-6}$ for the space and time increments. The
morphology presented in the following is only a part of the whole
system.

Presented in Fig. 2 is the periodic morphology of
phase-field-crystal variable $\psi(x,y)$ with the parameters:
$u=0.1$, $\varepsilon=-0.005$, and $\overline{\psi}=0$. The period
is equivalent to $2\pi$ within an error of $10^{-5}$. The center
of the white stripe corresponds the maximum $\psi_{\rm max}$ and
that of the black stripe the minimum $\psi_{\rm min}$. This
phase-field-crystal simulated dimerization pattern is in agreement
with experimental STM images of high-quality  Si(001) and Ge(001)
dimerized surfaces \cite{r001,ro,s-si,s-ge}. This stripe pattern
is obtained for the parameter region: $0.00221>\varepsilon >
-0.01$.

\begin{figure}[htb]
\begin{center}
\includegraphics[width=7.5cm,]{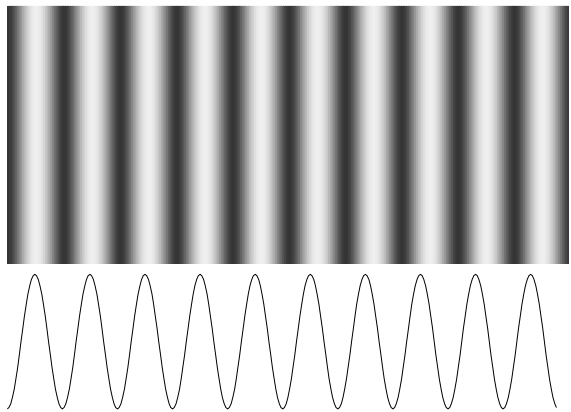}
\caption{Simulated striped morphology of phase-field-crystal
variable $\psi$ (upper part) and its side view curve (lower part).
We use the parameters $\delta x=\pi/16$, $\delta\tau=10^{-6}$,
$u=0.1$, $\varepsilon=-0.005$, and $\overline{\psi}=0$.}
\end{center}
\end{figure}

\begin{figure}[tbh]
\begin{center}
\includegraphics[width=8.2cm,]{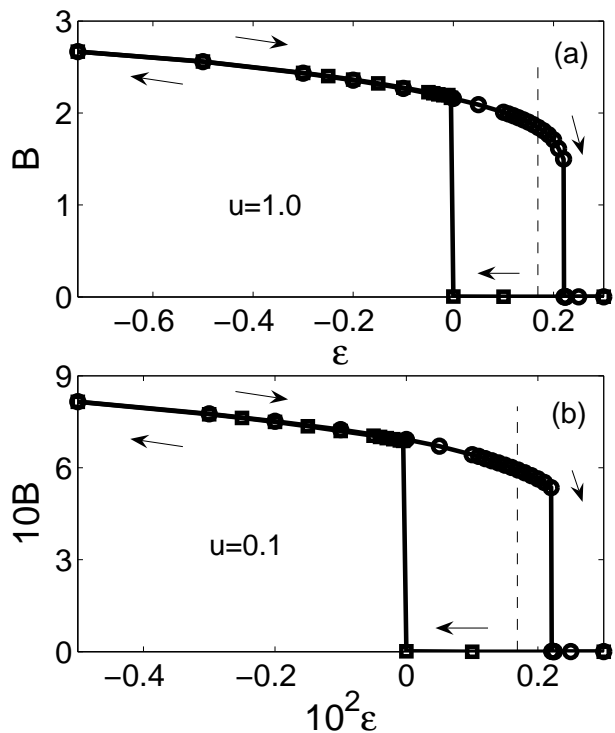}
\caption{The calculated parameter $B$ as a function of
$\varepsilon=a(T-T_L)$, with the phase-transition point
$\varepsilon_c$ indicated by dash line. The
$\varepsilon$-increasing part is shown by the solid line with
circles, and the $\varepsilon$-decreasing part by the solid line
with squares. The phase coexistence region is defined by
$\varepsilon_L\leq \varepsilon \leq \varepsilon_H$. The parameter
set $(\varepsilon_L,\varepsilon_c,\varepsilon_H)$, corresponding
to $(T_L,T_c,T_H)$, is $(0.0,0.169,0.221)$ for $u$=1.0, and
$(0.0,0.00169,0.00221)$ for $u$=0.1.  }
\end{center}
\end{figure}

We use $B=\psi_{\rm max}-\psi_{\rm min}$ as our order parameter
and present its $\varepsilon$ dependence for $u$=1.0 and $u$=0.1
in Fig. 3. It is clear that the phase transition is of first
order. The $B$, as a function of $\varepsilon$, in the case of
$u=0.1$ is ten times smaller than that in the case of $u=1.0$. The
$\varepsilon$ width of the phase coexistence region,
$\varepsilon_H-\varepsilon_L$, is proportional to $u^2$. Because
$\varepsilon$ is proportional to $T-T_L$, the $\varepsilon$
dependence implies the temperature dependence from zero
temperature to $T_L$, and finally beyond $T_H$. It is proved using
a series of simulated results that $T_c$ is also proportional to
$u^2$. Therefore, the parameter $u$ describes the temperature
width of phase coexistence region. Systematical analysis of
simulated results shows that $\psi$ can be quantitatively
described by the function
\begin{equation}
\psi=\frac B2\sin(\frac{q}{q_0}x),
\end{equation}
especially in the neighborhood of the maximum and the minimum.
Here $B$ and $q$ are determined by the simulated results.
$|q/q_0-1|$ can be very small as long as $u$ is small enough.
Actually, $q$ is equivalent to $q_0$ within a tiny error less than
10$^{-5}$ as long as $u$ is smaller than $0.1$.

On the other hand, $B$ and $q$ can be analytically determined by
minimizing the free energy (4) in terms of a variational
expression $\psi=C\sin(px)$ similar to Eq. (7). In this way, we
derive $p=1$ and
\begin{equation}
C=\frac 1{\sqrt{5}}\sqrt{3u+\sqrt{9u^2-40\alpha(T/T_L-1)}}.
\end{equation}
This means $q=q_0$. The expression (8) is reasonable only when $T$
is not larger than $T_H$, which implies that $u$ can be expressed
as
\begin{equation}
u=\frac 23 \sqrt{10\alpha(T_H-T_L)/T_L}.
\end{equation}
We can obtain an analytical expression of the order parameter
$B=2C$ for $u\leq 0.1$ by requiring that $C$ given by Eq. (8) is
equivalent to $B/2$ obtained by numerically solving Eq. (5). The
phase-transition temperature $T_c$ is given by
\begin{equation}
T_c=\frac 34T_H+\frac 14T_L.
\end{equation}
Using the relation $\phi=q_0\psi$ (due to $u_6=1$) and Eq. (9), we
obtain $B=\Delta b/c$ and
\begin{equation}
\Delta b=2c\sqrt[4]{\frac{
8\alpha(T_H-T_L)}{5T_L}}\sqrt{1+\sqrt{\frac{T_H-T}{T_H-T_L}}}
\end{equation}
for $T\leq T_H$, and $B=\Delta b=0$ for $T>T_H$. This expression
determines the temperature dependence of the dimerization-induced
change of the bond length, $\Delta b$.

\section{Applied to the (2$\times$1)-(1$\times$1) phase transitions of Si(001) and Ge(001)}

For the Si(001) and Ge(001)  dimerized surfaces, it is relatively
easy to measure $T_L$, $T_H$, $c_0$ (zero temperature), and
$\Delta b_{\rm RT}$ (room temperature, 300 K). It should be
noticed that $q=2\pi/\bar{c}$, where $\bar{c}=2c$
($\bar{c}_0=2c_0$). We use these parameters as input. The
parameter $\alpha$ can be determined by substituting 300 K and
$\Delta b_{\rm RT}$ for $T$ and $\Delta b$ in Eq. (11), and then
calculate $u$ and $T_c$ in terms of Eqs. (9) and (10). We
calculate $u_4$ using the definition $u_4=uq^2_0$ (due to
$u_6=1$). The input and calculated results are summarized in Table
I. In Fig. 4 we present the dimerization parameter $\Delta b$ as a
function of temperature for both the Si(001) and the Ge(001)
surfaces. It is clear that $\Delta b$ is still finite at $T_c$,
and jumps to zero once $T$ is larger than $T_H$.

\begin{table}[!htb]
\begin{center}
\caption{The input parameters and the calculated results for the
dimerized Si(001) and Ge(001) surfaces.}
\begin{tabular}{cccccccccccc}\hline \hline
  \multicolumn{3}{r}{Si(001)}
 & &   $T_L$ (K) & & $T_H$ (K) & & $c_0$ (\AA) & & $\Delta b_{\rm RT}$ (\AA) & \\
      &  & & &   1223   &  &  1473   &  &   3.84   &    &   1.6  & \\
\cline{5-11}  & &  &  &      $\alpha$ & & $u$   &   &   $u_4$    &  &   $T_c$ (K) & \\
     &  &  & &   5.8$\times$10$^{-4}$ & & 0.022   &   &   0.06  &  &   1410 & \\
\hline \multicolumn{3}{r}{Ge(001)}
   &  & $T_L$ (K) & & $T_H$ (K) & & $c_0$ (\AA) & & $\Delta b_{\rm RT}$ (\AA) & \\
     &   &  & &  950   & &  1130   & &   3.99   & &   1.6 & \\
\cline{5-11}   &  &  & &  $\alpha$ & & $u$   &   &   $u_4$   &   &   $T_c$ (K) & \\
     &  & & &  5.4$\times$10$^{-4}$ & &  0.020 & & 0.05  & & 1085 & \\
 \hline \hline
\end{tabular}
\end{center}
\end{table}

\begin{figure}[!tbh]
\begin{center}
\includegraphics[angle=0,origin=br,width=8.2cm,]{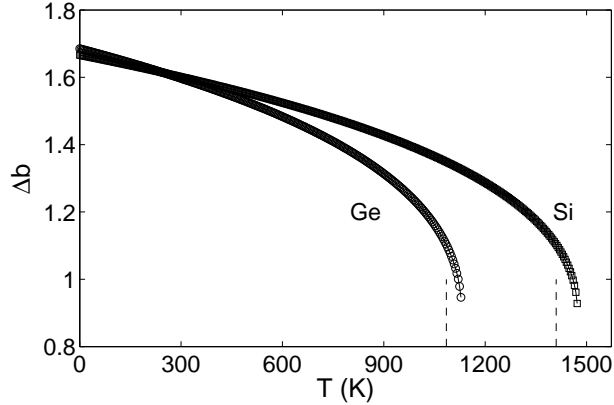}
\caption{The temperature dependences of dimerization parameter
$\Delta b$ (\AA) for the Si(001) (square) and Ge(001) (circle)
phases, with $T_c$ indicated by dash line. $T_H$ is 1473 K for the
Si(001) and 1130 K for the Ge(001). $\Delta b$ is finite at $T_c$,
and jumps to zero once $T>T_H$.  }
\end{center}
\end{figure}

It should be pointed out that our phase-field-crystal equilibrium
patterns, as shown in Fig. 2, consist of regular arrays of
infinitely long dimer chains. These are in agreement with
large-scale experimental STM images of the parallel perfect dimer
chains in the high-quality Si(001) and Ge(001) surfaces
\cite{s-si,s-ge}. Generally speaking, such a dimer chain can be
broken into several segments, but these segments are still
completely in the same line \cite{s-si,s-ge,r001,ro,cont,break}
and can be considered to be an infinitely long dimer chain in the
sense of averaging along the line. Actually, there is either
type-$S_A$ or type-$S_B$ dimerized phases in high-quality (001)
surfaces \cite{s-si,s-ge}. Therefore, our phase-field-crystal
theory can describe well the experimental regular arrays of dimer
chains.

In addition, our simulated results are in agreement with
experimental observation that the dimers still exist above $T_c$
\cite{cont}, as shown in  Fig. 4. Some of the dimers begin to
break at $T_L$ and all of them finally disappear at $T_H$. Between
$T_L$ and $T_H$, there is a phase coexistence of the normal dimers
and the broken ones, but there is no coexistence of the type-$S_A$
and type-$S_B$ discussed in earlier references \cite{mc}. This is
the clear sign of the first-order phase transition and thus there
must be some dimers above the phase transition temperature $T_c$.
Naturally, $T_H-T_c$ varies in different samples, and can be so
small that $T_H$ is equivalent to $T_c$ within measurement error,
which leads to some conclusions that there is no dimers above
$T_c$ \cite{break}. Essentially, a structural phase can be judged
only when it has a large enough spatial size. Therefore, our
modeling and simulated results are not only reasonable but also in
agreement with experiment \cite{cont,break}.

\section{Conclusion}

In summary, we have proposed a two-dimensional phase-field-crystal
model for the Si(001) and Ge(001) dimerized surface phases and
their phase transitions to corresponding $1\times 1$ phases at
elevated temperatures. We use a phase-field-crystal variable to
describe the dimerization of atoms in the top layer, and determine
it by solving the evolution equation derived from the free energy.
The simulated periodic arrays of dimer chains are consistent with
STM images of the Si(001) and Ge(001) dimerized surfaces. The
calculated temperature dependence of the dimerization parameter
shows that normal dimers and broken ones coexist between $T_L$ and
$T_H$ and the first-order structural phase transition takes place
at $T_c$ in between. These results are in agreement with
experiment. This phase-field-crystal approach can be directly
applied to phase transitions of semiconductor $n\times 1$
reconstructed surface phases, and should be suitable to other
semiconductor reconstructed surfaces and their phase transitions.

\section*{Acknowledgements}

This work is supported  by Nature Science Foundation of China
(Grant Nos. 10774180, 10874232, and 60621091), by Chinese
Department of Science and Technology (Grant No. 2005CB623602), and
by the Chinese Academy of Sciences (Grant No. KJCX2.YW.W09-5).

\end{document}